# Artificial Magnetic Conductor Frame to Improve Impedance Matching and Radiation Symmetry in 2×2 Array for 6G Applications

E. Giusti[(1,2)], K.K. Tiwari [(3)], C.J. Reddy [(4)], D. Brizi [(1,2)], A. Monorchio [(1,2)] and G. Caire[(3)]
(1) Department of Information Engineering, University of Pisa, Pisa, 56122, Italy (edoardo.giusti@phd.unipi.it, danilo.brizi@unipi.it, agostino.monorchio@unipi.it)
(2) Consorzio Nazionale Interuniversitario per le Telecomunicazioni (C.N.I.T.), Pisa, 56122, Italy
(3) Technical University of Berlin, Germany (tiwari@tu-berlin.de, caire@tu-berlin.de)
(4) Siemens Digital Industries Software, Troy, MI (cj.reddy@siemens.com)

*Abstract*— **We present an Artificial Magnetic Conductor (AMC) frame capable of improving the impedance matching and the radiation symmetry of a 2×2 dual polarized array without degrading isolation performance for 6G applications. The proposed frame is integrated into the array without modifying the single radiating element design. By relying on accurate full-wave simulations, it results that the addition of the frame restores the impedance matching performance, achieving a 1.5 GHz bandwidth in the 5G NR n257 band at 28 GHz. The isolation between each port remains under -15 dB within the operating band, thanks to the vias in the rectangular patch metasurface. Moreover, the overall structure exhibits a gain of 11.81 dBi with an aperture efficiency of 69%, satisfactorily for broadband communication purposes. The proposed AMC frame represents an effective method for improving array performance without the need to alter the shape or dimensions of the single radiating element.**

## I. Introduction

The development of 5G and the forthcoming advent of 6G communication standards require array systems operating at high frequencies with the motivations of broadband and ultra-high data rates [1], [2], [3]. In the literature, many patch antennas suitable for millimeter-wave array systems can be found. However, a major issue with this type of antenna relies in its integration into an array configuration, which often leads to a degradation of the radiating performance. One of the main problems encountered consists in the mutual coupling between array elements, which can be particularly critical when the elements exhibit strong resonance. For example, the directional mutual coupling of the respective orthogonal modes in the dual-polarized radiating element design of [4] causes asymmetric radiation pattern in the $2 \times 2$ 2x2 array configuration, which is unsuitable for the state-of-the-art array antenna architecture ([5], [6]), while the single radiating element itself has a symmetric radiation pattern.

A first and straightforward solution is to increase the substrate dimensions in order to physically separate the elements. However, changing the array proportions leads to a degradation of the overall impedance matching. To overcome this drawback, a properly designed Artificial Magnetic Conductor (AMC) frame can be employed. As it is well known from the literature, an ideal AMC is a metamaterial characterized by a reflection coefficient phase equal to 0°. For practical purposes, a metamaterial presenting a phase of the reflection coefficient within $-90°$ to $90°$ is operating as a good approximation of an AMC [7]. The integration of this frame allows an improvement in array performance in terms of impedance matching, without affecting inter-port isolation, and the radiation symmetry. Design simulations in this paper were done using commercially available EM simulation tool, Simcenter Feko (previously Altair Feko) [8].

The remainder of the paper is organized as follows. In Section II, the design and performance of the single antenna element is described, with particular focus on the AMC frame. Section III presents the results of the array configurations with and without the AMC. Finally, the conclusions are drawn.

## II. Single Element

The selected radiating element is a wideband dual-polarized patch antenna realized using the Printed Circuit Board (PCB) technology. The antenna proposed in [4] has been scaled to operate around a center frequency of 28 $GHz$. The complete antenna, including the AMC frame, is shown in Fig. 1. The antenna has a square shape with a side length $W_S = 7.089\ mm$ and a height of 2.5 $mm$. The substrate material is Taconic RF-35 ($\varepsilon_r = 3.5$, $\tan \delta = 0.0041$). With reference to Fig. 1, the remaining antenna dimensions are: $l_p = 3.115\ mm$, $w_p = 0.354\ mm$, $s = 0.177\ mm$, $g = 0.027\ mm$, $w_{pa} = 2.478\ mm$. The AMC frame has a width $w_{EBG} = 0.8\ mm$ and a length $l_{EBG} = 2.65\ mm$. Moreover, vias, with a diameter $v_{EBG} = 0.2\ mm$, are placed at the center of each rectangular element to mitigate surface currents. The overall frame extends the original patch by only 2 $mm$, resulting in an overall structure size of approximately $0.66\ \lambda$. The antenna is fed by two coaxial ports, enabling dual orthogonal polarizations. The feeding ports are positioned at a distance $l_f = 0.354\ mm$ from the center of the antenna.

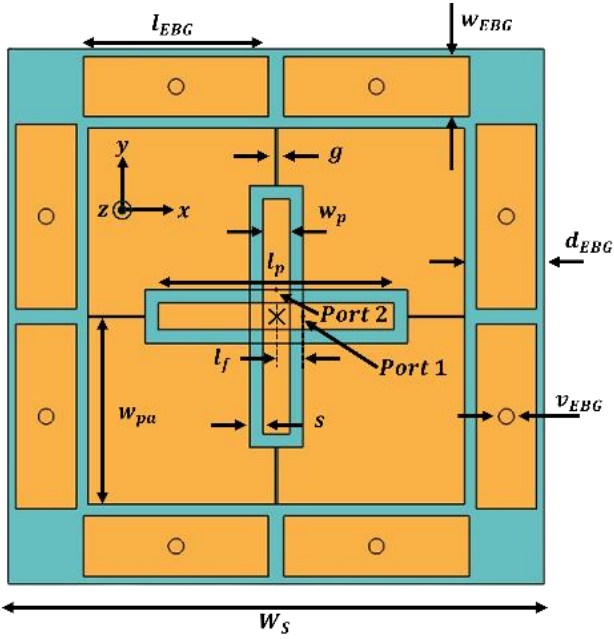

Figure 1. Front view of the proposed single array element with the integrated AMC frame. The substrate is shown in gray, while the metallic components are shown in orange.

To evaluate the performance of the single antenna element with the AMC frame, we first analyzed the S-parameters of the patch antenna without the AMC frame. As shown in Fig. 2, the antenna is poorly matched due to changes in its radiating behavior. Furthermore, the center frequency is shifted from the intended band center of $28\ GHz$. Although extending the substrate is mandatory to mitigate the inter-element mutual coupling, this extension alone is not sufficient to preserve proper impedance matching. Therefore, the AMC frame described previously was introduced. As observed in Fig. 3, when the AMC frame is added, without modifying the antenna dimensions, the magnitudes of $S_{11}$ and $S_{22}$ exhibit a bandwidth of $1.44\ GHz$ centered at $28\ GHz$. Moreover, as depicted in Fig. 4, the isolation between port 1 and port 2 is significantly improved, with both $S_{12}$ and $S_{21}$ magnitude remaining below $-15\ dB$ across the operating band. This confirms the effectiveness of the AMC frame in improving the matching performance while maintaining good isolation between the two orthogonal ports.

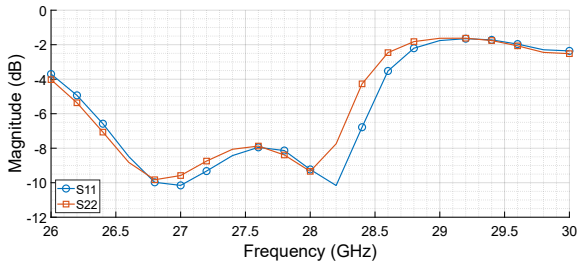

Figure 2. Magnitude of $S_{11}$ and $S_{22}$ for the single element patch antenna without the AMC frame.

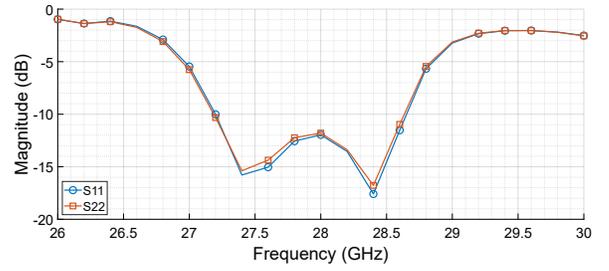

Figure 3. Magnitude of $S_{11}$ and $S_{22}$ for the single element patch antenna with the AMC frame.

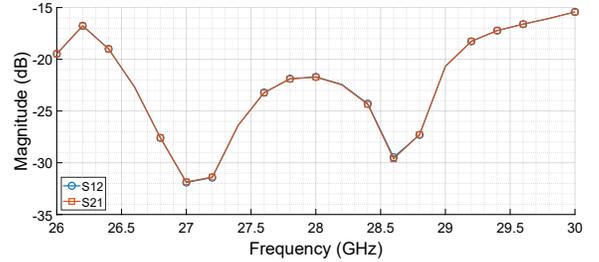

Figure 4. Magnitude of $S_{12}$ and $S_{21}$ for the single element patch antenna with the AMC frame.

## III. 2×2 ARRAY STRUCTURE AND RESULTS

### A. 2×2 Array without AMC Frame

To compare our approach with the conventional configuration in which array elements are simply placed adjacent to each other, a 2x2 array without the AMC frame was designed. As shown in Fig. 5, the substrate was extended in order to reduce mutual coupling between the elements with respect to the solution presented in [4]. The size of each single element is $0.66\lambda$, which is the same as the dimension of the patch antenna presented in the previous section. The array system has eight ports: the odd-numbered ports are dedicated to the polarization along the *x*-axis, while the even-numbered ports are used for the *y*-axis polarization. As observed in Fig. 6, this array configuration with an extended substrate exhibits very good performance with extremely reduced mutual coupling among the ports. The drawback lies in the degraded impedance matching of the individual elements. As shown in Fig. 7, all $S_{nn}$ parameters remain above the $-10\ dB$ threshold. This degradation is caused by the substrate extension, which alters the resonance of the structure. While this modification improves isolation between the elements, it does so at the expense of the impedance matching performance.

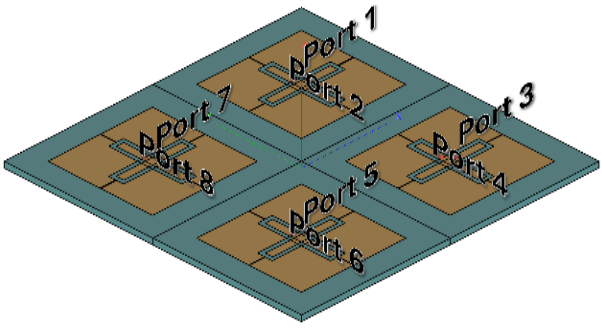

Figure 5. Isometric view of the 2x2 array configuration. Odd-numbered ports correspond to polarization along the *x*-axis, while even-numbered ports correspond to polarization along the *y*-axis.

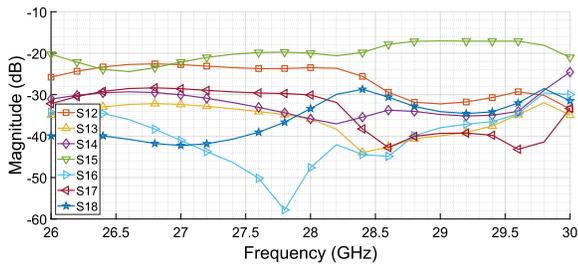

Figure 6. Isolation in terms of S-parameters between port 1 and the other seven ports of the array configuration without the AMC frame.

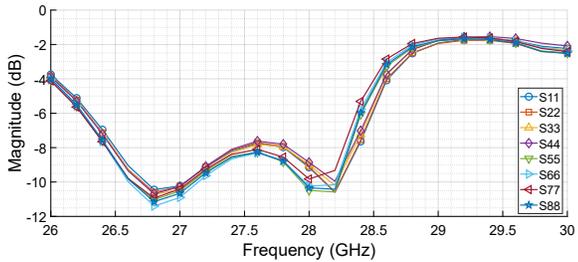

Figure 7. Magnitude of $S_{nn}$ for the array configuration without the AMC frame.

The extension of the antenna substrate not only degrades the matching performance of all eight ports, but also, modifies the radiation pattern. As shown in Fig. 8, the gain pattern of the 2x2 array, when the odd-numbered ports were fed, without the AMC frame is asymmetric, resulting in a Half-Power BeamWidths (HPBWs) of 34.83° and 38.58° for the $E_\theta$ and $E_\phi$ planes, respectively. This asymmetry is caused by directional inter-element mutual coupling, which is still present despite the extension of the substrate and the consequent increase in the inter-element spacing.

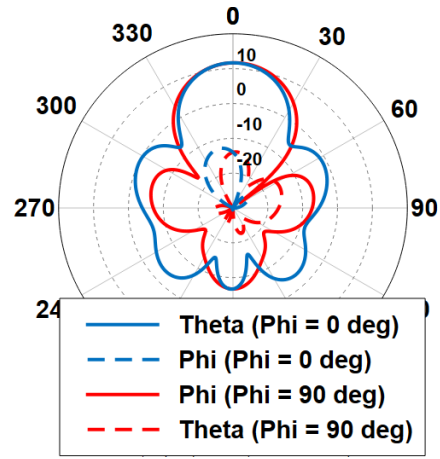

Figure 8. Gain polar plot of the 2x2 array without the AMC frame with the odd-numbered ports fed.

### B. 2×2 Array with AMC Frame

After examining the array composed solely of patch elements on an extended substrate, the array was evaluated using the patch antenna presented in Section II, which is equipped with the AMC frame. The port positions are the same as in the previous case, as shown in Fig. 2. As can be observed in Fig. 9, the impedance matching of the antenna is improved for all eight ports. This configuration exhibits an average bandwidth of 1.5 $GHz$, with similar response curves for each $S_{nn}$ parameter.

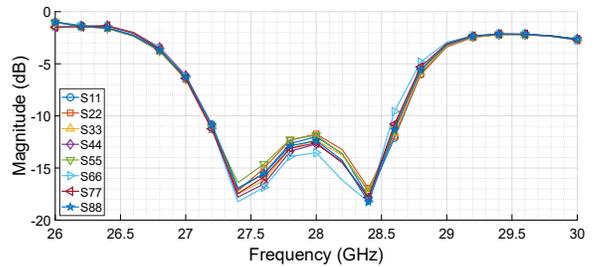

Figure 9. Magnitude of $S_{nn}$ for the array configuration with the AMC frame.

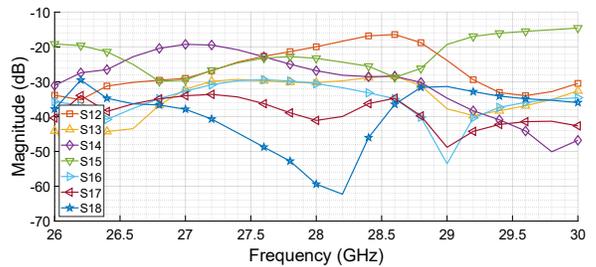

Figure 10. Isolation in terms of S-parameters between port 1 and the other seven ports of the array configuration with the AMC frame.

Fig. 10 shows the coupling between port 1 and the remaining seven ports. The mutual coupling remains below $-15\ dB$ within the operating band, thanks to the vias of the

AMC frame, which effectively mitigate the surface currents flowing between ports.

The gain pattern of the overall antenna is symmetric, as shown in Fig. 11. Thanks to its ability to mitigate surface waves, the AMC frame suppresses inter-element mutual coupling, thereby restoring the radiation pattern. The overall gain of the structure is 11.81 $dBi$, corresponding to an aperture efficiency of 69%. Moreover, the cross-polarization discrimination is 23.55 $dB$ and 22.14 $dB$ at 28 $GHz$ for the $\Phi = 0°$ and $\Phi = 90°$ planes, respectively. The HPBW is 40° for both the $E_\theta$ and $E_\phi$ main planes which is very critical for state-of-the-art beamforming performance [5].

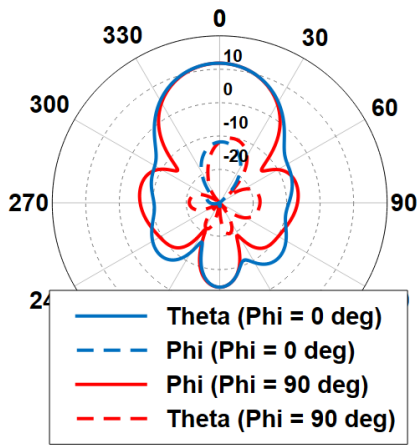

Figure 11. Gain polar plot of the 2×2 array with the AMC frame with the odd-numbered ports fed.

## IV. Conclusion

In this article, an AMC frame has been proposed to improve the impedance matching of an array system without degrading mutual coupling isolation. The frame, exhibiting AMC behavior, is designed to enhance impedance matching performance while simultaneously mitigating mutual coupling between individual array elements, aided by the presence of vias. The overall bandwidth of the array has been restored to 1.5 $GHz$, achieving an aperture efficiency of 69%. In conclusion, the proposed frame design represents an effective solution for addressing coupling issues in array configurations without requiring any reshaping of the single radiating element and thereby preserving all the merits of the stand-alone radiating element.